\documentclass[journal]{IEEEtran}
\usepackage{subfigure,amsmath}
\usepackage{booktabs} 
\usepackage{mathtools, cuted}
\usepackage{longtable}
\usepackage{graphicx}
\usepackage{algorithm}
\usepackage{comment}
\usepackage{algpseudocode}
\usepackage{amssymb}
\usepackage{setspace}
\usepackage{bbm}
\usepackage[pdftex,colorlinks,bookmarksopen,bookmarksnumbered,citecolor=red,urlcolor=red]{hyperref}
\usepackage{tikz}
\begin{document}



\title{An Alternative Data-Driven Prediction Approach Based on Real Option Theories}
\author{{ Abdullah AlShelahi, Jingxing Wang, Mingdi You, Eunshin Byon,~\IEEEmembership{Member,~IEEE}, and Romesh Saigal}

\thanks{Abdullah AlShelahi, Jingxing Wang, Eunshin Byon, and Romesh Saigal are with the Department of Industrial and Operations Engineering, University of Michigan, Ann Arbor, MI 48109. Mingdi You is with Ford Motor Company, 22001 Michigan Ave, Dearborn, MI 48124.  (corresponding author; shelahi@umich.edu). This work was supported by the National Science Foundation under Grant IIS-1741166.}}

\maketitle


\date{}

\begin{abstract}

This paper presents a new prediction model for time series data by integrating a time-varying Geometric Brownian Motion model with a pricing mechanism used in financial engineering. Typical time series models such as Auto-Regressive Integrated Moving Average assumes a linear correlation structure in time series data.  When a stochastic process is highly volatile, such an assumption can be easily violated, leading to inaccurate predictions. We develop a new prediction model that can flexibly characterize a time-varying volatile process without assuming linearity. We formulate the prediction problem as an optimization problem with unequal overestimation and underestimation costs. Based on real option theories developed in finance, we solve the optimization problem and obtain a predicted value, which can minimize the expected prediction cost. We evaluate the proposed approach using multiple datasets obtained from real-life applications including manufacturing, finance, and environment. The numerical results demonstrate that the proposed model shows competitive prediction capability, compared with alternative   approaches.
\end{abstract}

\section {Introduction} 
\label{sec-intro}

\IEEEPARstart{I}{n} many applications including manufacturing, energy, and finance, accurate prediction is required to support strategic, tactical and/or operational decisions of organization \cite{chatfield2000time}. When physical information about the underlying mechanism that generates the time series data is limited, data-driven methods can be useful for predicting future observations \cite{zhang2003time}. In general, data-driven forecasting methods predict future observations based on past observations \cite{box2015time}. Several data-driven methods have been proposed in the literature for modeling time series data, among which Auto-Regressive Integrated Moving Average (ARIMA) and its variants such as the ARIMA-General Auto Regressive Conditional Heteroskedasticity (ARIMA-GARCH) have been widely used in many applications due to their flexibility and statistical properties \cite{ruppertd,hahn2009electric,sohn2007hierarchical,lu2014portfolio}. ARIMA which assumes a constant standard deviation of stochastic noises, whereas  ARIMA-GARCH  extends it  by allowing the standard deviation to vary over time.

The typical ARIMA-based models estimate its model parameters using historical data and uses the estimated time-invariant parameters throughout the prediction period. Using such time-invariant parameters may not capture possible changes in the underlying data generation mechanism. Some studies modify the original ARIMA model to update the parameters using new observations \cite{tran2004automatic,ledolter1981recursive}. The basic idea of these ARIMA-based models is that the future observation can be predicted by using a linear combination of past observations (and estimated noises). Therefore they assume a linear correlation structure between consecutive observations  \cite{brooks2014introductory}. However, when the underlying dynamics exhibits a highly volatile process, such a simple linear structure may provide poor prediction performance \cite{kantz2004nonlinear}. 

This study aims to provide accurate predictions for a highly volatile and time-varying stochastic process whose underlying dynamics is complicated and possibly nonlinear. As an example, let us consider a prediction problem faced by a contract manufacturer (CM) located in Michigan in the U.S, which motivates this study.  The CM is a manufacturing company that produces various automotive parts, such as front and rear bumper beams, for several large automotive companies worldwide. The CM deals with a large number of orders for bumper beams from several automotive companies and the order sizes are time-varying. The CM should plan their production capacity carefully so that it can deliver products promptly when it gets orders. When an actual order size is greater than expected (i.e., when an order size is underestimated), overtime wages must be paid to workers to meet demands. On the other hand, when an order size is smaller than predicted (i.e., when an order size is overestimated), workers and equipment become idle. 

As such, CM wants to predict future order sizes accurately, so that it can reduce its operating costs resulting from the discrepancy between its predicted value and actual sizes. Currently, CM uses its own proprietary prediction model, but its prediction performance is not satisfactory. The details of CM's proprietary model are confidential, so we cannot find reasons for its unsatisfactory performance. When we apply the ARIMA and ARIMA-GARCH models to CM's datasets, we also do not obtain significantly better prediction results (detailed results will be provided in Section~ \ref{sec-case}).  We believe such poor performance of ARIMA-based approaches is because they cannot fully characterize the underlying volatile  dynamics. In addition to historical data, the future order size may depend on other factors which possibly make the order process behave nonlinearly. A new prediction approach that can adapt to such time-varying, and possibly nonlinear, dynamics is needed for providing better forecasts.

To this end we develop a new method for predicting future values in highly volatile processes, based on real option pricing theories typically used in financial engineering. One of the popularly used stochastic process models for pricing real options is the Geometric Brownian Motion (GBM) model. Brownian motion is a continuous-time stochastic process, describing random movements in time series variables. The GBM, which is a stochastic differential equation, incorporates the idea of Brownian motion and consists of two terms: a deterministic term to characterize the main trend over time and a stochastic term to account for random variations. In GBM the random variations are represented by Brownian Motion \cite{bjork2009arbitrage}. GBM is useful to model a positive quantity whose changes over equal and non-overlapping time intervals are identically distributed and independent. 

The GBM has been applied to represent various real processes in finance, physics, etc.  \cite{gardiner1986handbook}. In particular, it becomes a fundamental block for many asset pricing models \cite{bjork2009arbitrage}, and recently it has been applied to facilitate the use of a rich area of options theory to solve various pricing problems (see, for example, \cite{whitt1981stationary,thorsen1999afforestation,benninga2002real,nembhard2002real,boomsma2012renewable,chiu2017real}). However, most of the current GBM studies have been limited to solving pricing problems and have not used real options theory for making forecasts.  

In this study,  by utilizing the full power of real options theory, we present a new approach for predicting future observations when the system's underlying dynamics follows the GBM process.  Specifically, we allow the GBM parameters to adaptively change over time in order to characterize time-varying dynamics. We formulate the prediction problem as an optimization problem and provide a solution using real option theories. To the best of our knowledge, our study is the first attempt to incorporate options theory in the prediction problem. 

Our approach provides extra flexibility by allowing overestimation (or over-prediction) to be handled differently from underestimation (or over-prediction). The overestimation and underestimation costs are determined in real life applications, depending on a decision-maker's (or organization's) preference.  For example, in the aforementioned CM case, overestimation  and underestimation of order sizes could cause different costs. The CM may want to put a larger penalty on the demand underestimation than on the overestimation, so that it can avoid extra overtime wages. We incorporate unequal overestimation and underestimation costs into the optimization problem and find the optimal forecast that minimizes the expected prediction cost. 

To evaluate the prediction performance, we use three datasets collected from different applications, including the demand for bumper beams in CM (manufacturing), stock prices (finance), and wind speed (environment). We compare the performance of our model with ARIMA and ARIMA-GARCH models (and the proprietary prediction model in the CM case study) with different combinations of overestimation and underestimation costs. In most cases, our model outperforms those alternative models. In particular, we find that when the process is highly time-varying such as stock prices and wind speed, the proposed approach provides much stronger prediction capability than ARMA and ARIMA-GARCH. 
\raggedbottom 

The remainder of the paper is organized as follows. The mathematical formulation and solution procedure are discussed in Section~\ref{sec-meth}. Section~\ref{sec-case} provides numerical results in three different applications. Section~\ref{sec-con} concludes the paper.

\section {Methodology}\label{sec-meth}
\subsection {Problem Formulation}
Consider a real-valued  variable $S(t)$ which represents a system state at time $t$. For example, the state variable can be a stock market index price, a manufacturer's order size, or wind speed. This state variable is assumed to follow an inhomogeneous GBM with time-varying parameters. 

Let us consider a filtered probability space $(\Omega,\mathcal{F},P,\mathcal{F}_t)$, where the filtration $\mathcal{F}_t$ is generated by the Brownian motion $W$, i.e. $\mathcal{F}_t= \mathcal{F}_t^{{W}}$ so that $\mathcal{F}_t$ contains all information generated by ${W(t)}$, up to and including time $t$. With GBM, the stochastic process $S(t)$ is modeled by the following dynamics.
\begin{equation} \label{eqn-gbm}
dS(t)=\mu(t) S(t) dt + \sigma(t) S(t) dW(t),
\end{equation}
where $\sigma(t)$ denotes the  volatility of $S(t)$ and $\mu(t)$  represents a drift process. The stochastic process $W(t)$ represents the Brownian motion where the increment $W(t+\Delta t) - W(t)$ during the time interval $\Delta t$ is normally distributed  with mean 0 and variance $\Delta t$, denoted by $\mathcal{N}(0,\Delta t)$, and  $W(t)$ is assumed to be stationary. 

Our objective is to predict $S(T)$ in the future  time at $T (>t)$ when the current time is $t$.  Solving (\ref{eqn-gbm}) by using It\^{o}'s lemma \cite{shreve2004stochastic}, we obtain 
\begin{equation}\label{eqn-S}
S(T)=S(t)\: exp \Big( \int_t^T\Big(\mu(s)-\frac{1}{2}\sigma^2(s)\Big)ds +\int_t^T\sigma(s) dW(s)\Big),  
\end{equation}
and
\begin{equation} 
\mathbb{E}(S(T)|\mathcal{F}_t)= S(t)\:exp\big( \int_t^T\mu(s)ds \big). \label{eqn:exp_s}
\end{equation}

Let $K$ be the predicted  value of $S(T)$ at time $T$. When the overestimation and underestimation is penalized equally,  the quantity that represent the variable's central tendency, such as mean and median, is commonly used for prediction. But we consider a more general case where overestimation needs to penalized differently from underestimation, as discussed in Section~\ref{sec-intro}. When the observed value is $S(T)$, the overestimated quantity becomes  $\max\{K-S(T),0\}$, while the underestimated quantity is  $\max\{S(T)-K,0\}$. 

Let $p_o$ and $p_u$ denote the penalties for over/underestimation,  respectively. We formulate the  optimization problem for estimating $S(K)$ that can minimize the expected prediction cost, 
 \begin{eqnarray}
 \displaystyle \min_{K\in R^{+}} 
 \mathbb{E} \Big[P_o \max\{K-S(T),0\} + P_u \max\{S(T)-K,0\}|\mathcal{F}_t \Big].   \label{eqn-opt}
 \end{eqnarray}

Note that
\begin{equation}
\max\{K-S(T),0\}= K -S(T) + \max\{S(T)-K,0\}.\label{eqn-over}
\end{equation} 
If we substitute (\ref{eqn-over}) into (\ref{eqn-opt}), the optimal predicted value, denoted by $K^*$, can be obtained by solving the following objective function.

\begin{equation} 
\begin{split}
K(T)^* &= argmin_{K\in R^{+}} 
  \mathbb{E}  \Big[(P_o +P_u) \max\{S(T)-K,0\} \\& + P_o (K-S(T))|\mathcal{F}_t \Big],
  \end{split}
  \label{eqn-obj2}
\end{equation}
or equivalently, 

\begin{equation} 
\begin{split}
\displaystyle K(T)^* &= argmin_{K\in R^{+}}
\mathbb{E}  \Big[P_o \Big(\frac{P_o +P_u}{P_o} \max\{S(T)-K,0\} \\&+ (K-S(T))\Big)|\mathcal{F}_t \Big].
\end{split}
\label{eqn-obj3}
\end{equation}

In the next section we will present a solution procedure to obtain $K^*(T)$, based on the option theory. 

\subsection {Real Option Based Solution Procedure}
The optimization problem in (7) can be reformulated by employing the financial pricing theories. Suppose that we want to predict a state at the future time $T$. In pricing theories, $T$ can be viewed as the date to maturity, or the expiration date. 

A real option, also called contingent claim, with the date to maturity $T$, can be constructed on the state variable $S(t)$. A real option is a stochastic variable $\mathcal{X} \in \mathcal{F}_T^W$ that can be expressed as 
\begin{equation}
\mathcal{X}= \Phi(S(T)),
\end{equation}
where $\Phi(\cdot)$ is a contract function.

The contract function $\Phi(\cdot)$ is typically set to the payoff of the real option at time $T$. When the predicted value is $K$, $K$ can be viewed as the strike value in the option theory, while $\max\{S(T)-K,0\}$ is the payoff. Therefore, we get 
\begin{equation}
\Phi(S(T)) =\max\{S(T)-K,0\}
\end{equation}
It is required that $\mathcal{X} \in \mathcal{F}_T^W$ ensures that the value of the payoff of the real option $\mathcal{X}$ is determined at time $T$. 

Let the price process $\Pi(t;\mathcal{X})$ for the real option at time $t$ be given by a function $F(t,S(t)) \in [t,T] \times R_+$, i.e., 
\begin{equation}
\Pi(t;\mathcal{X})= F(t,S(t)).
\end{equation}
Here $F(\cdot)$ is a function which is assumed to be once continuously differentiable in $t$, and twice in $S(t)$. 

For a short-term prediction, the time interval $\Delta t$ between the current time $t$ and the future time $T$ is small, so we can assume that $\mu(t)$ and $\sigma(t)$ are constants during $[t,T]$. Then $F(t,S(t))$ can be obtained  by solving the Black-Scholes Partial Differential Equation (PDE) \cite{shreve2004stochastic},
 \begin{equation}
 \begin{split}
 &\frac{\partial{F(t,S(t))}}{\partial{t}} + \mu(t) S(t) \frac{\partial{F(t,S(t))}}{\partial{S}} \\ & + \frac{1}{2} S(t)^2 \sigma^2(t) ) \frac{\partial{F(t,S(t))}}{\partial{S^2}} - rF(t,S(t)) =0
  \end{split}
 \label{eqn-PDE1}
   \end{equation}
   with
    \begin{eqnarray}
  F(T,S(T))&=\Phi(S(T)), \label{eqn-PDE2}
   \end{eqnarray}
where $r$ represents a discounting factor.

 The Black-Scholes PDE in (\ref{eqn-PDE1})-(\ref{eqn-PDE2}) is usually solved numerically. But alternatively, we solve it using the Feyman-Ka$\breve{c}$ stochastic representation formula \cite{shreve2004stochastic}, to obtain 
\begin{eqnarray}
F(t,S(t))=e^{-r\Delta t} \mathbb{E}_{S}\big[\Phi(S(T))\ \mid \mathcal{F}_t \big]. \label{optionprice}
\end{eqnarray}

Next, we derive $F(t,S(t))$ in a closed form, given $K$. Letting $y= ln\Big [{S(T)}/{S(t)} \Big]$ and using the fact that $S(T)=S(t) exp((\mu(t) - \frac{1}{2} \sigma^2(t)) \Delta t + \sigma(t) \Delta W(t))$, it follows that $y \sim \mathcal{N}\big( \big(\mu(t)-\frac{1}{2}\sigma^2(t) \big)\Delta t, \sigma^2(t) \Delta t\big)$.
Thus, the probability density function $f(y)$ of $y$ is given by
\begin{eqnarray}
f(y)= \frac{1}{\sigma(t)\sqrt{2 \pi \Delta t}} e^-{\left( \frac{(y- (\mu(t)-\frac{1}{2} \sigma^2(t)) \Delta t)^2}{2 \sigma(t)^2 \Delta t}\right)}.
\end{eqnarray}
Consequently, we obtain

\begin{align}
\displaystyle
&\mathbb{E}_{S}\big[\Phi(S(T))\ \mid \mathcal{F}_t \big] \nonumber \\
& =\mathbb{E}_{S}\big[ \max\{S(T)-K,0\}|\mathcal{F}_t \big] \\  
&= \mathbb{E}_{S}\big[ \max\{S(t) e^y-K,0\} \big] \\ 
& =\int_{ln\frac{K}{S(t)}}^{\infty} S(t) e^y f(y) dy  - \int_{ln\frac{K}{S(t)}}^{\infty} K f(y) dy \label{eqn-exp} 
\end{align}

To solve (\ref{eqn-exp}), let $I_1$ and $I_2$, respectively, denote the first and second terms in (\ref{eqn-exp}).  We also let  $z= {y- (\mu(t)-0.5 \sigma^2(t)) \Delta t}/{\sigma(t)\sqrt{\Delta t}}$. First, $I_2$ becomes 
\begin{align}
I_2  =& \int_{ln\frac{K}{S(t)}}^{\infty} K f(y) dy \\  =&  K  \int_{-d2}^{\infty} \frac{1}{\sqrt{2 \pi}}  e^{\frac{- z^2}{2}}dz \\  =&  K  \int_{-\infty}^{d2} \frac{1}{\sqrt{2 \pi}}  e^{\frac{- z^2}{2}}dz= K \mathcal{N}(d_2)
\end{align}
where $d_2= {ln(\frac{S(t)}{K}) + (\mu(t) - \frac{1}{2} \sigma^2 )\Delta t}/{\sigma(t)\sqrt{\Delta t}}$ and $\mathcal{N}(\cdot)$ denotes the cumulative distribution function (CFD) for the standard normal distribution.  
Next, we obtain $I_1$ as 
\begin{align}
I_1  =&  \int_{ln\frac{K}{S(t)}}^{\infty} S(t) e^y f(y) dy \\ =& S(t) \int_{-d2}^{\infty} \frac{1}{\sqrt{2 \pi}}  e^{\frac{- z^2}{2} + z \sigma(t)\sqrt{\Delta t} + ( \mu(t) - \frac{1}{2} \sigma^2(t)) \Delta t}dz \\  =& S(t) \int_{-d2}^{\infty} \frac{1}{\sqrt{2 \pi}}  e^{\frac{-1}{2} ( z -\sigma(t)\sqrt{\Delta t} )^2 } e^{(\mu(t) \Delta t)}dz \\
 =&  S(t) e^{\mu(t) \Delta t} \int_{-d_2 - \sigma \sqrt{\Delta t}}^{\infty}  \frac{1}{\sqrt{2 \pi}}  e^{\frac{- v^2}{2}}dv \label{eqn-i1-chg} \\
 =& S(t) e^{\mu(t) \Delta t}  \mathcal{N}(d_1)  \label{eqn-d1}
\end{align}
where we use $v= z -\sigma(t)\sqrt{\Delta t} $ in (\ref{eqn-i1-chg}) and $d_1 = d_2+\sigma(t) \sqrt{\Delta t}$ in (\ref{eqn-d1}).

For small $\Delta t$, we can set $r=0$. 
Then, $F(t,S(t))$ in (\ref{optionprice}) becomes:
\begin{equation} \label{eq-F}
F(t,S(t))= e^{\mu(t) \Delta t} \: \mathcal{N}(d_1)S(t) - \mathcal{N}(d_2) \:K.
\end{equation} 
Note that given $\mu(t)$ and $S(t)$ at the current time $t$ and $K$, we can obtain $F(t,S(t))$. 

With the obtained expected payoff $\mathbb{E}_{S}\big[\Phi(S(T))\ \mid \mathcal{F}_t \big]$ where $\Phi(S(T)) =\max\{S(T)-K,0\}$,
we can find the optimal $K^*(T)$ in (\ref{eqn-obj3}). Let $\omega$ denote the ratio of overestimation cost to underestimation cost, i.e., 
\begin{align}  \label{eqn-w}
\omega=\frac{P_u}{P_o}
\end{align}
Given the price of the real option, defined in (\ref{optionprice}), we can reformulate the optimization problem in (\ref{eqn-obj3}) as
\newpage
\begin{strip}
\begin{align}
 K^*(T) =& \: argmin_{K\in R^{+}} \: \mathbb{E}  \left[(1+\omega) \max\{S(T)-K,0\} \mid \mathcal{F}_t  \right] + \mathbb{E} \left[(K-S(T)) \mid \mathcal{F}_t  \right] \label{eqn-obj5}  \\
 =& argmin_{K\in R^{+}}
\Big[(1+\omega) F(t,S(t)) + K-S(t) \: e^{\mu(t) \Delta t} \Big] \label{eqn-obj4} \\
=& argmin_{K\in R^{+}} 
\Big[(1+\omega) \left(e^{\mu(t) \Delta t} \: \mathcal{N}(d_1)S(t) - \mathcal{N}(d_2) \:K \right) + K-S(t) \: e^{\mu(t) \Delta t} \Big] \label{eqn-obj6} 
\end{align} 
\end{strip}
\noindent with  $d_2= {ln(\frac{S(t)}{K}) + (\mu(t) - \frac{1}{2} \sigma^2 )\Delta t}/{\sigma(t)\sqrt{\Delta t}}$ and $d_1 = d_2+\sigma(t) \sqrt{\Delta t}$. We use (\ref{optionprice}) with $r=0$ in the first term in the second equality and the last term in the second equality is obtained using (\ref{eqn:exp_s}). By plugging $F(t,S(t))$ in (\ref{eq-F}), we get the last equality.  

The predictor $K^*$ prefers overestimation when $\omega>1$ or underestimation when $\omega <1$. When overestimation and underestimation are equally penalized, the optimal $K^*$  can be obtained with $w=1$ in (\ref{eqn-obj5}).    
The optimization function in (\ref{eqn-obj5}) is a convex optimization problem that can be solved efficiently by existing numerical optimization softwares. In our implementation, we use Scipy's (Scientific Python) optimization library in Python.

\subsection {Parameters Estimation}\label{Parameters}

For a volatile stochastic process, the parameters $\mu(t)$ and $\sigma(t)$ can be time-varying. We estimate the nonstationary parameters using recent observations. Consider $n$ recent observations at the current time $t$, i.e., $S(t-(n-1)\Delta t) , S(t-(n-2)\Delta t), \cdots, S(t)$.  
Because $S(t)$ follows geometric Brownian motion and $\mu(t)$ and $\sigma(t)$ are assumed to be constant during the short interval $\Delta t$, the discretization scheme of (\ref{eqn-S})  is given by
\setlength{\arraycolsep}{0.0em}
\begin{equation}
\begin{split}
\ln\left(\frac{S(t+\Delta t)}{S(t)}\right) &= \left(\mu(t)-\frac{1}{2}\sigma^2(t)\right)\Delta t \\ &+\sigma(t) \big(W(t+\Delta t)- W(t) \big).  
\end{split}
\label{discretizationscheme}
\end{equation}

Noting that under GBM $\ln\Big({S(t+\Delta t)}/{S(t)}\Big)$ is normally distributed with mean $\big[\mu(t)-\frac{1}{2}\sigma^2(t)\big]\Delta t $ and variance $\sigma^2(t)$, we estimate  $\mu(t)$ and $\sigma(t)$ using maximum likelihood method as 
\begin{equation}
\begin{split}
\hat{\sigma}(t) =& \Bigg(\frac{1}{n}  \sum_{i=2}^{n} \Big(\ln\Big(\frac{S(t-(n-i)\Delta t)}{S(t-(n-i+1)\Delta t)}\Big)  \\ -& \frac{1}{n}\sum_{i=1}^{n-1}\Big[\ln\Big(\frac{S(t-(n-i)\Delta t)}{S(t-(n-i+1)\Delta t)}\Big)\Big]\Big)^2 \Bigg)^{\frac{1}{2}},
\end{split}
\label{eq-32}
\end{equation}
\begin{equation} 
\begin{split}
 \hat{\mu}(t) &=\frac{1}{n}\sum_{i=1}^{n}\Big[\ln\Big(\frac{S(t-(n-i)\Delta t}{S(t-(n-i+1)\Delta t)}\Big)\Big] \\ &+ \frac{1}{2} \hat{\sigma}(t)^2, 
 \end{split}
 \label{eq-33}
\end{equation}
respectively.

The estimated parameters $\hat{\mu}(t)$ and $\hat{\sigma}(t)$ are plugged into (\ref{eq-F}) and we obtain the optimal predicted value $K^*$ for $S(t+\Delta t)$ by solving~\eqref{eqn-obj6}.

\subsection {Implementation Details} \label{sec-imp-detail}

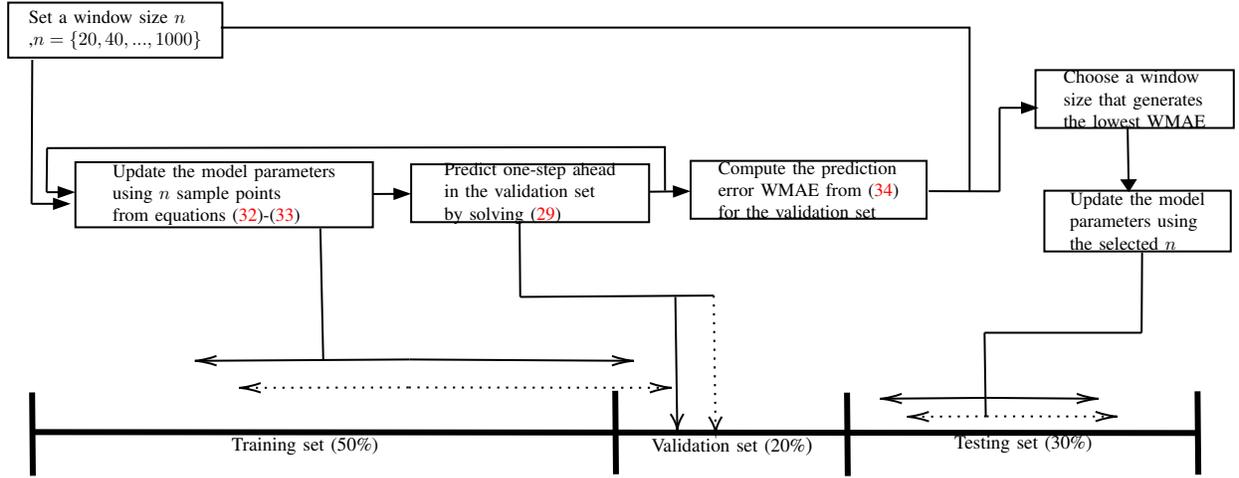
\begin{figure*}
\centering
\tikzset{every picture/.style={line width=0.75pt}} 

\begin{tikzpicture}[x=0.75pt,y=0.75pt,yscale=-0.73,xscale=1]
\draw [line width=2.25]    (25.75,301.59) -- (613.12,302.21) ;

\draw [line width=2.25]    (25.14,274.16) -- (25.52,331.83) ;

\draw [line width=2.25]    (319.24,273.06) -- (319.62,330.74) ;

\draw [line width=2.25]    (612.92,273.38) -- (613.31,331.05) ;

\draw [line width=2.25]    (436.14,274.32) -- (436.52,332) ;

\draw    (577.64,90.21) -- (578.22,133.73) ;
\draw [shift={(578.25,135.73)}, rotate = 269.23] [fill={rgb, 255:red, 0; green, 0; blue, 0 }  ][line width=0.75]  [draw opacity=0] (8.93,-4.29) -- (0,0) -- (8.93,4.29) -- cycle    ;

\draw   (120.3,21.92) -- (497.95,21.92) -- (497.95,135.08) ;
\draw    (477.36,135.01) -- (513.08,135.15) ;

\draw    (513.08,135.15) -- (513.08,77.38) ;

\draw    (513.08,77.38) -- (530.83,77.38) ;
\draw [shift={(532.83,77.38)}, rotate = 180] [fill={rgb, 255:red, 0; green, 0; blue, 0 }  ][line width=0.75]  [draw opacity=0] (8.93,-4.29) -- (0,0) -- (8.93,4.29) -- cycle    ;

\draw    (24.87,143.76) -- (25.23,44.24) ;

\draw    (24.87,143.76) -- (42.62,143.76) ;
\draw [shift={(44.62,143.76)}, rotate = 180] [fill={rgb, 255:red, 0; green, 0; blue, 0 }  ][line width=0.75]  [draw opacity=0] (8.93,-4.29) -- (0,0) -- (8.93,4.29) -- cycle    ;

\draw   (32.56,105.25) -- (344.28,104.54) -- (344.36,134.6) ;
\draw    (32.45,135.73) -- (32.71,105.27) ;

\draw    (32.45,135.73) -- (44.94,135.73) ;
\draw [shift={(46.94,135.73)}, rotate = 180] [fill={rgb, 255:red, 0; green, 0; blue, 0 }  ][line width=0.75]  [draw opacity=0] (8.93,-4.29) -- (0,0) -- (8.93,4.29) -- cycle    ;

\draw    (271,155.74) -- (271.32,208.14) ;

\draw    (215.5,251.39) -- (326.53,252.33) ;
\draw [shift={(328.53,252.34)}, rotate = 180.48] [color={rgb, 255:red, 0; green, 0; blue, 0 }  ][line width=0.75]    (10.93,-3.29) .. controls (6.95,-1.4) and (3.31,-0.3) .. (0,0) .. controls (3.31,0.3) and (6.95,1.4) .. (10.93,3.29)   ;

\draw    (215.5,251.39) -- (109.24,252.33) ;
\draw [shift={(107.24,252.34)}, rotate = 359.49] [color={rgb, 255:red, 0; green, 0; blue, 0 }  ][line width=0.75]    (10.93,-3.29) .. controls (6.95,-1.4) and (3.31,-0.3) .. (0,0) .. controls (3.31,0.3) and (6.95,1.4) .. (10.93,3.29)   ;

\draw  [dash pattern={on 0.84pt off 2.51pt}]  (215.03,270.73) -- (345.64,270.96) ;
\draw [shift={(347.64,270.96)}, rotate = 180.1] [color={rgb, 255:red, 0; green, 0; blue, 0 }  ][line width=0.75]    (10.93,-3.29) .. controls (6.95,-1.4) and (3.31,-0.3) .. (0,0) .. controls (3.31,0.3) and (6.95,1.4) .. (10.93,3.29)   ;

\draw  [dash pattern={on 0.84pt off 2.51pt}]  (215.03,270.73) -- (131.5,270.99) ;
\draw [shift={(129.5,271)}, rotate = 359.82] [color={rgb, 255:red, 0; green, 0; blue, 0 }  ][line width=0.75]    (10.93,-3.29) .. controls (6.95,-1.4) and (3.31,-0.3) .. (0,0) .. controls (3.31,0.3) and (6.95,1.4) .. (10.93,3.29)   ;

\draw    (506.68,277.91) -- (560.9,278.51) ;
\draw [shift={(562.9,278.54)}, rotate = 180.64] [color={rgb, 255:red, 0; green, 0; blue, 0 }  ][line width=0.75]    (10.93,-3.29) .. controls (6.95,-1.4) and (3.31,-0.3) .. (0,0) .. controls (3.31,0.3) and (6.95,1.4) .. (10.93,3.29)   ;

\draw    (506.68,277.91) -- (454.83,278.51) ;
\draw [shift={(452.83,278.54)}, rotate = 359.33000000000004] [color={rgb, 255:red, 0; green, 0; blue, 0 }  ][line width=0.75]    (10.93,-3.29) .. controls (6.95,-1.4) and (3.31,-0.3) .. (0,0) .. controls (3.31,0.3) and (6.95,1.4) .. (10.93,3.29)   ;

\draw  [dash pattern={on 0.84pt off 2.51pt}]  (506.45,290.7) -- (570.4,290.84) ;
\draw [shift={(572.4,290.85)}, rotate = 180.13] [color={rgb, 255:red, 0; green, 0; blue, 0 }  ][line width=0.75]    (10.93,-3.29) .. controls (6.95,-1.4) and (3.31,-0.3) .. (0,0) .. controls (3.31,0.3) and (6.95,1.4) .. (10.93,3.29)   ;

\draw  [dash pattern={on 0.84pt off 2.51pt}]  (506.45,290.7) -- (468.5,290.98) ;
\draw [shift={(466.5,291)}, rotate = 359.57] [color={rgb, 255:red, 0; green, 0; blue, 0 }  ][line width=0.75]    (10.93,-3.29) .. controls (6.95,-1.4) and (3.31,-0.3) .. (0,0) .. controls (3.31,0.3) and (6.95,1.4) .. (10.93,3.29)   ;

\draw    (584.96,177.52) -- (584.7,232.91) ;

\draw    (584.7,232.91) -- (505.49,232.91) ;

\draw    (506.11,290.7) -- (505.49,232.91) ;

\draw  [dash pattern={on 0.84pt off 2.51pt}]  (368.89,207.19) -- (369.84,299.9) ;
\draw [shift={(369.86,301.9)}, rotate = 269.42] [color={rgb, 255:red, 0; green, 0; blue, 0 }  ][line width=0.75]    (10.93,-3.29) .. controls (6.95,-1.4) and (3.31,-0.3) .. (0,0) .. controls (3.31,0.3) and (6.95,1.4) .. (10.93,3.29)   ;

\draw    (349.57,208.14) -- (350.51,298) ;
\draw [shift={(350.54,300)}, rotate = 269.4] [color={rgb, 255:red, 0; green, 0; blue, 0 }  ][line width=0.75]    (10.93,-3.29) .. controls (6.95,-1.4) and (3.31,-0.3) .. (0,0) .. controls (3.31,0.3) and (6.95,1.4) .. (10.93,3.29)   ;

\draw    (368.89,207.19) -- (271.32,208.14) ;

\draw    (170.53,160.76) -- (171.98,251.93) ;

\draw    (13.1,4.68) -- (121.1,4.68) -- (121.1,42.68) -- (13.1,42.68) -- cycle  ;
\draw (67.1,23.68) node [scale=0.7] [align=left] {Set a window size $n$\\,$n= \{20,40,...,1000\}$};
\draw    (46.99,115) -- (196.99,115) -- (196.99,160) -- (46.99,160) -- cycle  ;

\draw (121.99,137.5) node [scale=0.7] [align=left]  {Update the model parameters \\ using $n$ sample points\\ from equations (\ref{eq-32})-(\ref{eq-33})};
\draw    (216.32,116.44) -- (336.32,116.44) -- (336.32,156.44) -- (216.32,156.44) -- cycle  ;
\draw (276.32,136.44) node [scale=0.7] [align=left] {Predict one-step ahead \\in the validation set \\by solving (\ref{eqn-obj4})};
\draw    (357.27,114.07) -- (476.27,114.07) -- (476.27,154.07) -- (357.27,154.07) -- cycle  ;
\draw (416.77,134.07) node [scale=0.7] [align=left] {Compute the prediction \\error WMAE from (\ref{eq-34})\\for the validation set};
\draw    (531.2,51.32) -- (631.2,51.32) -- (631.2,91.32) -- (531.2,91.32) -- cycle  ;
\draw (581.2,71.32) node [scale=0.7] [align=left]  {Choose a window \\size that generates \\the lowest WMAE};
\draw    (535.67,134.69) -- (629.67,134.69) -- (629.67,176.69) -- (535.67,176.69) -- cycle  ;
\draw (582.67,155.69) node [scale=0.7] [align=left] {Update the model \\parameters using \\the selected $n$};
\draw (162.8,311.05) node [scale=0.8] [align=left]  {{\small Training set (50\%)}};
\draw (378.23,311.05) node [scale=0.8] [align=left] {{\small Validation set (20\%})};
\draw (525.07,310.11) node [scale=0.8] [align=left]  {{\small Testing set (30\%)}};
\draw    (336.32,135.42) -- (355.27,135.1) ;
\draw [shift={(357.27,135.07)}, rotate = 539.03] [fill={rgb, 255:red, 0; green, 0; blue, 0 }  ][line width=0.75]  [draw opacity=0] (8.93,-4.29) -- (0,0) -- (8.93,4.29) -- cycle    ;

\draw    (196.99,136.98) -- (214.32,136.87) ;
\draw [shift={(216.32,136.85)}, rotate = 539.6] [fill={rgb, 255:red, 0; green, 0; blue, 0 }  ][line width=0.75]  [draw opacity=0] (8.93,-4.29) -- (0,0) -- (8.93,4.29) -- cycle    ;

\end{tikzpicture}

  \caption{Overall procedure of the proposed approach (the dotted lines imply that the model parameters are updated in a rolling-horizon manner using the most $n$ recent observations)}
  \label{f1}
\end{figure*}

We refer our proposed model to as the \textit{option prediction model}.  Figure~\ref{f1} summarizes the overall procedure of the proposed approach. We also summarize the procedure of the proposed approach in Algorithm~1 below. We set the time step $\Delta t =1$ to make the one-ahead step prediction. The data is divided into three sets: training, validation, and testing. The training set starts at $t=1$ and ends at $t=N_1$, consisting of about 50\% of the entire data set, is used to determine the model parameters as shown in Figure \ref{f1}. The validation set,  consisting of about 20\% of the data set, is used for determining the window size $n$.  Lastly the testing set consists of the  last $30 \%$ of the data set and it starts at $t=N_2$.

\begin{algorithm}
\caption{Option prediction model}\label{algo}
\begin{algorithmic}[1]
\State \textbf{Initialization:}
\State Choose a window size $n$ by validation as shown in Figure \ref{f1}.
\State Obtain initial estimates for the model parameters $\hat{\sigma}(N_2)$ and  $\hat{\mu}(N_2)$ in (\ref{eq-32}) and (\ref{eq-33}), respectively.
\State Determine $F(N_2,S(N_2))$ in  (\ref{optionprice}).
\For{$k=N_{2}+1$ to $\infty$}
\State \textbf{Prediction:}
\State Obtain $K^*$ by solving (\ref{eqn-obj6}) to obtain the one-step ahead state prediction.
\State \textbf{Update:}
\State Observe $S(k)$.
\State Obtain $\hat{\sigma}(k)$ and  $\hat{\mu}(k)$ in (\ref{eq-32}) and (\ref{eq-33}), respectively, by using $n$ recent observations.
\State Determine $F(k,S(k))$ in  (\ref{optionprice}).
\EndFor
\end{algorithmic}
\end{algorithm}

In Algorithm~1 we determine the window size $n$ for  obtaining the parameters $\hat{\mu}(t)$ and $\hat{\sigma}(t)$, we use the validation technique \cite{friedman2001elements}.  We fit the model with a different window size $n$ and evaluate the prediction performance using  data in the validation set and choose the best window size that generates the lowest prediction error in the validation set. The performance of our approach is evaluated using data in the testing set (See Figure \ref{f1}). We report the prediction performance in the testing set in 
Section~\ref{sec-case}.

In evaluating the prediction performance, we 
consider that the overestimated and underestimated prediction results need to be evaluated differently for $\omega\neq 1$. As such we employ  the following two performance measures, namely,  Weighted Mean Absolute Error (WMAE) and Weighted Mean Absolute Percentage Error (WMAPE), defined by
\begin{equation}
\begin{split}
\textrm{WMAE} &= \frac{1}{N} \sum_{t=1}^{N}\Big( \mathbbm{1}_{(S(t)>K^*(t))} \omega |S(t) - K^*(t)| \\
 &+   \mathbbm{1}_{(S(t)<K^*(t)} |S(t) - K^*(t)| \Big) 
 \end{split}
\label{eq-34}
\end{equation}

and 
\begin{equation}
\begin{split}
\textrm{WMAPE} &= \frac{1}{N} \sum_{t=1}^{N} \Big(\frac{  \mathbbm{1}_{(S(t)>K^*(t))} \omega |S(t) - K^*(t)|} {S(t)}\\
& + \frac{\mathbbm{1}_{(S(t)<K^*(t))} |S(t) - K^*(t)|}{S(t)} \Big),
\end{split}
 \end{equation}
respectively, where $N$ denotes the number of data points in the testing set and $K^*(t)$ is the predicted value at time $t$.

\section {Case Studies} \label{sec-case}
 This section implements the proposed prediction model using multiple datasets obtained from real-life applications. Specifically we examine the performance of the predictive model in predicting the size of a manufacturer's order,  a stock market index price, and  wind speed.

\subsection {Alternative methods}

We compare our model with two standard time series models, namely, the ARIMA and the ARIMA-GARCH. We use the Akaika Information Criteria (AIC) to select the model order in  both models. For fair comparison, we update the model parameters in a rolling horizon manner, similar to the procedure discussed in Section~\ref{sec-imp-detail}. That is, we determine the window size $n$ using the validation technique and update the model parameters using the most recent $n$ observations whenever a new observation is obtained. 
 
 With underestimation penalties, Pourhab et al. \cite{pourhabib2015short} suggest using quantile of the predictive state density. With $\omega (=p_u/p_o)$ denoting the ratio of underestimation cost to overestimation cost, we use the $({\omega}/{1+\omega})$-quantile, given by 
\begin{equation} \label{adjust}
\textrm{Quantile prediction}=\hat{\mu}_a(t) + \hat{\sigma}_a(t) \Phi ^{-1} \Big(\frac{\omega}{1+\omega}\Big),
\end{equation}
where $\hat{\mu}_a(t)$ denotes the estimated predicted mean, $\hat{\sigma}_a$ is the estimated standard deviation in ARIMA (or ARIMA-GARCH) model,   and $ \Phi^{-1} (\cdot)$ denotes the inverse of the standard  normal CDF. Note that large (small) $w$ puts more penalty on $p_u$ ($p_o$) and the quantile prediction provides a larger (smaller) predicted value, so underestimation (overestimation) can be avoided. 

\subsection {Manufacturing Data}
We first study the prediction problem faced by our industry partner,  CM. The historical data obtained from CM includes orders of 10 different types of bumper beams. We use monthly data on those 10 types of bumper beams ordered over a period of 29 consecutive months (the order size varies from 0 to over 36,000 items). When applying the proposed model to this problem, the choice of weight $\omega$ affects the final prediction. By changing the weight, we are able to show a preference for over-capacity (overestimation) or under-capacity (underestimation). We consider different cases for choosing the weight parameter $\omega$. 

Let us first look at the case when $\omega$ is set to be less than one (i.e, $p_u\le p_o$). According to CM, workers and equipment can be shifted from one type of bumper beam to another, but doing so incurs 15$\%$ loss of production efficiency. In other words, if one type of bumper beam is overestimated, causing over-capacity,  available resources can be assigned to other bumper beam production, but with a reduced efficiency. 
In this case, underestimation  is favored and we set $w=1/1.15$.

Next the weight parameter can be  set to be greater than 1 (i.e, $p_u\ge p_o$) when the prediction is preferred to be more than the actual order size. According to the labor law in Michigan in the U.S., overtime rate is higher than the regular salary. In this case we set $w=1.15$ to emphasize the preference of overestimation to underestimation. Finally we also consider $w=1$, which reflects equal penalties. 

The errors in terms of WMAE and WMAPE for all ten types of bumper beams  are presented in Tables~\ref{tbl-cm-w1}-\ref{tbl-cm-w3} with three different weights. Overall our option prediction model performs better than the CM's own prediction, ARIMA and ARIMA-GARCH in both criteria. With $w=1/1.15$ the proposed approach provides lower WMAEs (WMAPEs) for 9 (5) types of bumper beams out of 10 types. Similarly, with other $w$ values, our approach outperforms the alternative models in most cases. 

\begin{table}[H]
 	\small
 	\centering
 	\caption{CM Prediction Results for ten types of bumper beams with $\omega=1/1.15$ in the  Testing Set (The values in bold indicate the lowest prediction error for each product)}\label{tbl-cm-w1}
 	
 	\resizebox{!}{.08\paperheight}{
 		\begin{tabular}{c c c c c}
 			\hline \hline
 			\multicolumn{5}{c}{\textbf{Weighted Mean Absolute Error (WMAE)}} \\
            \hline                  
            {\textbf{Product No.}}   & \textbf{ARIMA} & \textbf{Option Prediction} & \textbf{ARIMA-GARCH} & \textbf{CM Prediction} \\
            \hline \hline
    1     & 1196.58 & \textbf{411.97} & 2105.25 & 1161.56 \\
    \midrule
    2     & 332.36 & \textbf{127.90} & 168.72 & 151.24 \\
    \midrule
    3     & 119.35 & \textbf{105.49} & 107.43 & 194.69  \\
    \midrule
    4     & 1476.09 & \textbf{574.52} & 2185.30 & 936.94  \\
    \midrule
    5     & 1330.40 & \textbf{1299.00} & 1327.08 & 1797.74   \\
    \midrule
    6     & 542.33 & 64.38 & 63.24 & \textbf{42.90} \\
    \midrule
    7     & 357.38 & \textbf{24.14} & 497.54 & 92.17 \\
    \midrule
    8     & 1776.17 & \textbf{1339.11} & 3103.77 & 2520.75  \\
    \midrule
    9     & 1496.62 & \textbf{1305.49} & 2887.48 & 2475.71 \\
    \midrule
    10    & 1125.52 & \textbf{516.06} & 3278.77 & 1928.58 \\
\vspace{0.01in}
		\end{tabular}}

	\resizebox{!}{.08\paperheight}{
 		\begin{tabular}{c c c c c}
 			\hline \hline
 			\multicolumn{5}{c}{\textbf{Weighted Mean Absolute Percent Error (WMAPE)}} \\
            \hline                  
            {\textbf{Product No.}}   & \textbf{ARIMA} & \textbf{Option Prediction} & \textbf{ARIMA-GARCH} & \textbf{CM Prediction} \\

            \hline \hline
    1     & 1.14  & \textbf{0.19} & 0.94  & 0.52 \\
    \midrule
    2     & 0.62  & 0.43  & 0.42  & \textbf{0.40} \\
    \midrule
    3     & 28.89 & 0.44  & \textbf{0.41} & 0.69 \\
    \midrule
    4     & 1.20  & 0.69  & 3.92  & \textbf{0.66} \\
    \midrule
    5    & 0.12  & \textbf{0.11} & \textbf{0.12} & 0.15 \\
    \midrule
    6     & 45.23 & \textbf{5.30} & 11.01 & 7.93 \\
    \midrule
    7     & 215.17 & \textbf{3.66} & 408.98 & 9.63 \\
    \midrule
    8    & \textbf{0.24} & 0.26  & 0.76  & 0.60 \\
    \midrule
    9    & \textbf{0.19} & 0.22  & 0.66  & 0.58 \\
    \midrule
    10    & 0.36  & \textbf{0.20} & 2.63  & 0.99 \\
\bottomrule
\bottomrule
		\end{tabular}}
 \end{table}
 
 \begin{table}[H]
 	\small
 	\centering
 	\caption{CM Prediction Results for ten types of bumper beams with $\omega=1$ in the  Testing Set (The values in bold indicate the lowest prediction error for each product)}\label{tbl-cm-w2}
 	
 	\resizebox{!}{.08\paperheight}{
 		\begin{tabular}{c c c c c}
 			\hline \hline
 		\multicolumn{5}{c}{\textbf{Weighted Mean Absolute Error (WMAE)}} \\
            \hline                  
            {\textbf{Product No.}}   & \textbf{ARIMA} & \textbf{Option Prediction} & \textbf{ARIMA-GARCH} & \textbf{CM Prediction} \\
            \hline \hline
    1     & 1193.29 & \textbf{537.94} & 2103.43 & 1335.79 \\
    \midrule
    2     & 328.39 & \textbf{134.67} & 164.49 & 168.93\\
    \midrule
    3     & 111.78 & 118.39 & \textbf{103.40} & 223.89\\
    \midrule
    4     & 1307.50 & \textbf{609.06} & 1920.83 & 1073.16 \\
    \midrule
    5     & 1178.90 & 1403.17 & \textbf{1200.81} & 2060.95 \\
    \midrule
    6     & 472.86 & 74.52 & 57.08 & \textbf{45.12}\\
    \midrule
    7     & 315.42 & \textbf{34.61} & 434.33 & 93.51 \\
    \midrule
    8     & 1691.28 & \textbf{1465.94} & 2737.90 & 2590.93 \\
    \midrule
    9     & 1453.32 & \textbf{1397.14} & 2564.50 & 2527.94 \\
    \midrule
    10    & 1008.59 & \textbf{569.58} & 2865.30 & 1932.47 \\
\vspace{0.01in}
		\end{tabular}}

	\resizebox{!}{.08\paperheight}{
 		\begin{tabular}{c c c c c}
 			\hline \hline
 			\multicolumn{5}{c}{\textbf{Weighted Mean Absolute Percent Error (WMAPE)}} \\
            \hline                  
            {\textbf{Product No.}}   & \textbf{ARIMA} & \textbf{Option Prediction} & \textbf{ARIMA-GARCH} & \textbf{CM Prediction} \\
   
            \hline \hline
    1     & 1.09  & \textbf{0.24} & 0.94  & 0.59 \\
    \midrule
    2    & 0.61  & 0.45  & \textbf{0.40} & 0.44 \\
    \midrule
    3   & 25.76 & 0.50  & \textbf{0.39} & 0.79 \\
    \midrule
    4    & 1.05  & \textbf{0.72} & 3.41  & 0.75 \\
    \midrule
    5     & 0.11  & 0.12  & \textbf{0.11} & 0.17 \\
    \midrule
    6   & 39.61 & \textbf{6.17} & 9.59  & 7.95 \\
    \midrule
    7   & 187.36 & \textbf{3.87} & 356.23 & 9.67 \\
    \midrule
    8      & \textbf{0.22} & 0.28  & 0.66  & 0.60 \\
    \midrule
    9    & \textbf{0.18} & 0.23  & 0.58  & 0.58 \\
    \midrule
    10   & 0.31  & \textbf{0.21} & 2.29  & 0.99 \\
\bottomrule
\bottomrule
		\end{tabular}}
 \end{table}
 \vspace{-0.09in}
 \begin{table}[H]
 	\small
 	\centering
 	\caption{CM Prediction Results for ten  types of bumper beams with $\omega=1.15$  in the  Testing Set (The values in bold indicate the lowest prediction error for each product)}\label{tbl-cm-w3}
 	\resizebox{!}{.08\paperheight}{
 		\begin{tabular}{c c c c c}
 			\hline \hline
 			 \multicolumn{5}{c}{\textbf{Weighted Mean Absolute Error (WMAE)}} \\
            \hline                  
            {\textbf{Product No.}}   & \textbf{ARIMA} & \textbf{Option Prediction} & \textbf{ARIMA-GARCH} & \textbf{CM Prediction} \\
            \hline \hline
    1     & 1196.58 & \textbf{411.97} & 2105.25 & 1161.56 \\
    \midrule
    2     & 332.36 & \textbf{127.90} & 168.72 & 151.24  \\
    \midrule
    3     & 119.35 & \textbf{105.49} & 107.43 & 194.69\\
    \midrule
    4     & 1476.09 & \textbf{574.52} & 2185.30 & 936.94 \\
    \midrule
    5     & 1330.40 & \textbf{1299.00} & 1327.08 & 1797.74 \\
    \midrule
    6     & 542.33 & 64.38 & 63.24 & \textbf{42.90}  \\
    \midrule
    7     & 357.38 & \textbf{24.14} & 497.54 & 92.17  \\
    \midrule
    8     & 1776.17 & \textbf{1339.11} & 3103.77 & 2520.75 \\
    \midrule
    9     & 1496.62 & \textbf{1305.49} & 2887.48 & 2475.71 \\
    \midrule
    10    & 1125.52 & \textbf{516.06} & 3278.77 & 1928.58 \\
\vspace{0.01in}
		\end{tabular}}

 	\resizebox{!}{.08\paperheight}{
 		\begin{tabular}{c c c c c}
 			\hline \hline
 			\multicolumn{5}{c}{\textbf{Weighted Mean Absolute Percent Error (WMAPE)}} \\
            \hline                  
            {\textbf{Product No.}}   & \textbf{ARIMA} & \textbf{Option Prediction} & \textbf{ARIMA-GARCH} & \textbf{CM Prediction} \\
            \hline \hline
    1     & 1.14  & \textbf{0.19} & 0.94  & 0.52 \\
    \midrule
    2    & 0.62  & 0.43  & 0.42  & \textbf{0.40} \\
    \midrule
    3   & 28.89 & 0.44  & \textbf{0.41} & 0.69 \\
    \midrule
    4     & 1.20  & 0.69  & 3.92  & \textbf{0.66} \\
    \midrule
    5    & 0.12  & \textbf{0.11} & 0.12 & 0.15 \\
    \midrule
    6     & 45.23 & \textbf{5.30} & 11.01 & 7.93 \\
    \midrule
    7     & 215.17 & \textbf{3.66} & 408.98 & 9.63 \\
    \midrule
    8    & \textbf{0.24} & 0.26  & 0.76  & 0.60 \\
    \midrule
    9    & \textbf{0.19} & 0.22  & 0.66  & 0.58 \\
    \midrule
    10   & 0.36  & \textbf{0.20} & 2.63  & 0.99 \\
\bottomrule
\bottomrule
		\end{tabular}}
 \end{table}
Although ARIMA and ARIMA-GARCH provide the lowest errors for  some products,  their prediction performance is not consistent. For example, for $1^{st}$, $4^{th}$ and $7^{th}$ product, WMAEs from ARMA are much higher than the proposed approach, whereas ARIMA-GARCH results in pretty poor performance for predicting order sizes for $8^{th}-10^{th}$ products. On the contrary, our approach provides more stable results. Even when WMAEs and WMAPEs from our approach are higher than other approaches, they  are close to the lowest errors. Therefore, we can conclude that our approach is more accurate and reliable. The CM’s proprietary model does not account for unequal weights on overestimation and underestimation. If the company wants to  minimize the excess inventory due to overestimation, a small (less than 1) weight parameter should be assigned. If the company goal is to meet customer satisfaction, overestimation should be preferred with a large (larger than 1) weight parameter. In this sense our approach can reflect the company's management preference more flexibly.

\subsection {Stock Market Index Data}

To evaluate the performance of our approach in a highly volatile process, we consider stock market index price time series data. We analyze the daily closing price of the Dow Jones index in three time periods between 2010 and 2015. 

Risk averse and risk seeking investors have different preferences in terms of overestimation and underestimation. That being said, in a bull market, stock prices are expected to increase. In such a case, risk seeking investors with aggressive investment strategies would prefer biasing their prediction to overestimation. On the contrary, risk averse investors tend to be less optimistic, making them conservative, preferring underestimation. To reflect different investment preferences, we consider three values of the weight parameter $\omega$,  $1/1.15$, $1$, or $1.15$, to represent the underestimation preference, neutral/no preference, and overestimation preference, respectively. 

Table~\ref{tbl-stock} summarizes the results with three testing periods. Each testing period includes 100 days. Clearly, our option prediction performs better than ARIMA-GARCH and ARIMA in all cases, alerting for the possibility of a profitable trading strategy. The ARMA and ARMA-GARCH models generate 2.5 to 10 times higher WMAEs and 2 to 11 times higher WMAPEs.

    \begin{table}[H]
    \centering
      \caption{Dow Jones Index Price Prediction Results in the  Testing Set (The values in bold indicate the lowest prediction error for each testing period and weight)} \label{tbl-stock}
      \resizebox{!}{.17\paperheight}{
        \begin{tabular}{rclcc}
        \toprule
        \multicolumn{1}{c}{\textbf{Testing Period }} & \textbf{Weight} ($\omega$) & \multicolumn{1}{c}{\textbf{Method }} & \textbf{WMAE} & \textbf{WMAPE} \\
        \midrule
        \midrule
                       &                &  ARIMA-GARCH            &    595.54	& 0.04998 \\
                       & 1/1.15         & Option Prediction & \textbf{50.40} & \textbf{0.0043} \\
                       &                & ARIMA          &        628.69 &	0.0615   \\
\cmidrule{2-5}                       &                & ARIMA-GARCH            &   594.58	& 0.0499 \\
        \multicolumn{1}{l}{\textbf{Oct 2010- Mar 2011}} & 1              & Option Prediction & \textbf{54.30} & \textbf{0.0046} \\
                       &                & ARIMA          &    553.05	& 0.0541 \\
\cmidrule{2-5}                       &                & ARIMA-GARCH             & 682.72	& 0.0572 \\
                       & 1.15           & Option Prediction &    \textbf{58.78}     & \textbf{0.0050} \\
                       &                & ARIMA          &    559.59	& 0.0548 \\
        \midrule
                       &         & ARIMA-GARCH          &    360.39	& 0.0237 \\
                       &    1/1.15             & Option Prediction & \textbf{70.94} & \textbf{0.0046} \\
                       &                & ARIMA          &       455.35	& 0.0309  \\
\cmidrule{2-5}                       &                &ARIMA-GARCH             & 317.34	& 0.0209 \\
        \multicolumn{1}{l}{\textbf{Aug 2013 - Dec 2013}} & 1              & Option Prediction &    \textbf{75.78}     &  \textbf{0.0049}\\
                       &                & ARIMA          &  406.97	& 0.0276 \\
\cmidrule{2-5}                       &                & ARIMA-GARCH             &   321.71	& 0.0211\\
                       & 1.15           & Option Prediction & \textbf{81.09} & \textbf{0.0052} \\
                       &                & ARIMA          &     418.90	& 0.0284 \\
        \midrule
                       &                & ARIMA-GARCH  & 338.71 & 0.0195  \\
                       & 1/1.15         & Option Prediction &      \textbf{97.78}    &\textbf{0.0056}  \\
                       &                & ARIMA          &       182.60	& 0.0109 \\
\cmidrule{2-5}                       &                & ARIMA-GARCH             & 319.72	& 0.0184 \\
        \multicolumn{1}{l}{\textbf{Oct 2014 - Mar 2015}} & 1              & Option Prediction & \textbf{104.70} & \textbf{0.0060} \\
                       &                & ARIMA          &    166.65 &	0.0099 \\
\cmidrule{2-5}                       &                & ARIMA-GARCH             &     348.25	& 0.0200 \\
                       & 1.15           & Option Prediction & \textbf{113.11} & \textbf{0.0065} \\
                       &                & ARIMA          &   175.25	& 0.0104 \\
        \bottomrule
        \bottomrule
        \end{tabular}}
    \end{table}%

\subsection {Wind Speed Data}

Finally, we additionally consider another highly volatile process, wind speed. Because of environmental considerations, wind power, as a renewable source of energy, has been increasingly adopted worldwide \cite{byon2015adaptive}. Intermittent output of the farm is considered a challenging issue in terms of integrating the wind power into electric power grids. For reliable supply of power, steady and uninterrupted energy generation is desirable, which is not the case with wind energy. Wind speed is highly variable, depending on weather conditions and geographical factors such as the terrain. Such variability imposes challenges in power grid operations. To overcome the challenges, accurate forecasting of wind speed is required \cite{soman2010review}.

We use wind speed data collected from a meteorological tower near a wind farm  located in Europe.  The whole dataset consists of about  3000 samples, which covers a period of about a month. 
Due to the data confidentiality required by our industry partner, we omit more detailed description of the dataset studied in this case study. In wind farm operations some operators want to put a higher penalty on overestimation to avoid unsatisfied demand (or unsatisfied commitment), whereas underestimating wind speeds may be preferred when the salvage cost of excessively generated power is high \cite{pourhabib2015short,HG2010_PCE}. To reflect different costs, we use three different values for $\omega$, $1/1.15$, $1$ and $1.15$. 

Table~\ref{tbl-wind} summarizes the prediction results in the testing set from the three models. The proposed option prediction significantly outperform the other methods. The WMAEs and WMAPEs from ARMA and ARIMA-GARCH are higher by one order of magnitude than our approach. It demonstrates the superior prediction performance of  our approach in a  highly volatile process. 

\begin{table}[H]
 
		\centering
		\caption{Wind Speed Prediction Results in the Testing Set (The values in bold indicate the lowest prediction error for each weight)}		\label{tbl-wind} 
        \scalebox{0.7}{
        \begin{tabular}{clcc} 
        \toprule
        \textbf{Weight ($\omega$)} & \multicolumn{1}{c}{\textbf{Method }} & \textbf{WMAE} & \textbf{WMAPE} \\
        \midrule
        \midrule
                       & ARIMA-GARCH       &  3.364	& 0.402
   \\
        1/1.15         & Option Prediction & \textbf{0.318	
}  & \textbf{0.040} \\
                       & ARIMA          &8.73	&0.957
\\
        \midrule
                       & ARIMA-GARCH          &3.31	&0.380
   \\
        1              & Option Prediction & \textbf{0.342}  & \textbf{0.043} \\
                       & ARIMA          &8.73	& 0.96
 \\
        \midrule
                       & ARIMA-GARCH            & 	 3.74	
      & 0.422\\
        1.15           & Option Prediction & \textbf{0.365	
}  & \textbf{0.046} \\
                       & ARIMA          & 10.03	
	
          &1.100\\
        \bottomrule
        \bottomrule
        \end{tabular}}

\end{table}

\section{Conclusion}\label{sec-con}

In this study, we present a new prediction methodology for the time series data, based on option theories in finance when the underlying dynamics is assumed to follow the GBM process.  To characterize time-varying patterns, we allow the GBM model parameters to vary over time and update the parameter values using recent observations. We formulate the prediction problem with unequal overestimation and underestimation penalties as the stochastic optimization problem and provide its solution procedure.  We demonstrate the prediction capability of the proposed approach in various applications. Our approach appears to work well in the manufacturing application, when the order size varies over time.  For more highly volatile processes such as stock prices and wind speeds, the proposed model exhibits  much stronger prediction capability, compared to alternative ARIMA-based models. 

In the future, we plan to investigate other parameter updating schemes.  In this study, we update parameters in a rolling horizon manner using the maximum likelihood estimations. Another possibility is to use the Kalman filtering or its variants. Long-term predictions are beyond the scope of this study, but we plan to extend the approach presented in this study for obtaining accurate long-term predictions. We will also incorporate prediction results into managerial decision-making in several applications such as power grid operation with renewable energy \cite{Bouffard2008}.



\bibliographystyle{IEEEtran}
\bibliography{Ref_pricing_model.bib}

\begin{thebibliography}{10}
\providecommand{\url}[1]{#1}
\csname url@samestyle\endcsname
\providecommand{\newblock}{\relax}
\providecommand{\bibinfo}[2]{#2}
\providecommand{\BIBentrySTDinterwordspacing}{\spaceskip=0pt\relax}
\providecommand{\BIBentryALTinterwordstretchfactor}{4}
\providecommand{\BIBentryALTinterwordspacing}{\spaceskip=\fontdimen2\font plus
\BIBentryALTinterwordstretchfactor\fontdimen3\font minus
  \fontdimen4\font\relax}
\providecommand{\BIBforeignlanguage}[2]{{%
\expandafter\ifx\csname l@#1\endcsname\relax
\typeout{** WARNING: IEEEtran.bst: No hyphenation pattern has been}%
\typeout{** loaded for the language `#1'. Using the pattern for}%
\typeout{** the default language instead.}%
\else
\language=\csname l@#1\endcsname
\fi
#2}}
\providecommand{\BIBdecl}{\relax}
\BIBdecl

\bibitem{chatfield2000time}
C.~Chatfield, \emph{Time-series forecasting}.\hskip 1em plus 0.5em minus
  0.4em\relax CRC Press, 2000.

\bibitem{zhang2003time}
G.~P. Zhang, ``Time series forecasting using a hybrid {ARIMA} and neural
  network model,'' \emph{Neurocomputing}, vol.~50, pp. 159--175, 2003.

\bibitem{box2015time}
G.~E. Box, G.~M. Jenkins, G.~C. Reinsel, and G.~M. Ljung, \emph{Time series
  analysis: forecasting and control}.\hskip 1em plus 0.5em minus 0.4em\relax
  John Wiley \& Sons, 2015, pp. 1--2.

\bibitem{ruppertd}
D.~Ruppert, \emph{Statistics and Data Analysis for Financial
  Engineering}.\hskip 1em plus 0.5em minus 0.4em\relax Springer Texts in
  Statistics, 2015.

\bibitem{hahn2009electric}
H.~Hahn, S.~Meyer-Nieberg, and S.~Pickl, ``Electric load forecasting methods:
  Tools for decision making,'' \emph{European journal of operational research},
  vol. 199, no.~3, pp. 902--907, 2009.

\bibitem{sohn2007hierarchical}
S.~Y. Sohn and M.~Lim, ``Hierarchical forecasting based on {AR-GARCH} model in
  a coherent structure,'' \emph{European Journal of Operational Research}, vol.
  176, no.~2, pp. 1033--1040, 2007.

\bibitem{lu2014portfolio}
X.~F. Lu, K.~K. Lai, and L.~Liang, ``Portfolio value-at-risk estimation in
  energy futures markets with time-varying copula-garch model,'' \emph{Annals
  of operations research}, vol. 219, no.~1, pp. 333--357, 2014.

\bibitem{tran2004automatic}
N.~Tran and D.~A. Reed, ``Automatic {ARIMA} time series modeling for adaptive
  {I/O} prefetching,'' \emph{IEEE Transactions on parallel and distributed
  systems}, vol.~15, no.~4, pp. 362--377, 2004.

\bibitem{ledolter1981recursive}
J.~Ledolter, ``Recursive estimation and adaptive forecasting in {ARIMA} models
  with time varying coefficients,'' in \emph{Applied Time Series Analysis
  II}.\hskip 1em plus 0.5em minus 0.4em\relax Elsevier, 1981, pp. 449--471.

\bibitem{brooks2014introductory}
C.~Brooks, \emph{Introductory econometrics for finance}.\hskip 1em plus 0.5em
  minus 0.4em\relax Cambridge university press, 2002.

\bibitem{kantz2004nonlinear}
H.~Kantz and T.~Schreiber, \emph{Nonlinear time series analysis}.\hskip 1em
  plus 0.5em minus 0.4em\relax Cambridge university press, 2004, vol.~7.

\bibitem{bjork2009arbitrage}
T.~Bj{\"o}rk, \emph{Arbitrage theory in continuous time}.\hskip 1em plus 0.5em
  minus 0.4em\relax Oxford university press, 2009.

\bibitem{gardiner1986handbook}
C.~Gardiner, ``Handbook of stochastic methods for physics, chemistry and the
  natural sciences,'' \emph{Applied Optics}, vol.~25, p. 3145, 1986.

\bibitem{whitt1981stationary}
W.~Whitt, ``The stationary distribution of a stochastic clearing process,''
  \emph{Operations Research}, vol.~29, no.~2, pp. 294--308, 1981.

\bibitem{thorsen1999afforestation}
B.~J. Thorsen, ``Afforestation as a real option: Some policy implications,''
  \emph{Forest Science}, vol.~45, no.~2, pp. 171--178, 1999.

\bibitem{benninga2002real}
S.~Benninga and E.~Tolkowsky, ``Real options-- an introduction and an
  application to {R\&D} valuation,'' \emph{The Engineering Economist}, vol.~47,
  no.~2, pp. 151--168, 2002.

\bibitem{nembhard2002real}
H.~B. Nembhard, L.~Shi, and M.~Aktan, ``A real options design for quality
  control charts,'' \emph{The engineering economist}, vol.~47, no.~1, pp.
  28--59, 2002.

\bibitem{boomsma2012renewable}
T.~K. Boomsma, N.~Meade, and S.-E. Fleten, ``Renewable energy investments under
  different support schemes: A real options approach,'' \emph{European Journal
  of Operational Research}, vol. 220, no.~1, pp. 225--237, 2012.

\bibitem{chiu2017real}
C.-H. Chiu, S.-H. Hou, X.~Li, and W.~Liu, ``Real options approach for
  fashionable and perishable products using stock loan with regime switching,''
  \emph{Annals of Operations Research}, vol. 257, no. 1-2, pp. 357--377, 2017.

\bibitem{shreve2004stochastic}
S.~E. Shreve, \emph{Stochastic calculus for finance II: Continuous-time
  models}.\hskip 1em plus 0.5em minus 0.4em\relax Springer Science \& Business
  Media, 2004, vol.~11.

\bibitem{friedman2001elements}
J.~Friedman, T.~Hastie, and R.~Tibshirani, \emph{The elements of statistical
  learning: data mining, inference, and prediction}, ser. Springer Series in
  Statistics.\hskip 1em plus 0.5em minus 0.4em\relax Springer, 2nd ed., 2009.

\bibitem{pourhabib2015short}
A.~Pourhabib, J.~Z. Huang, and Y.~Ding, ``Short-term wind speed forecast using
  measurements from multiple turbines in a wind farm,'' \emph{Technometrics},
  vol.~58, no.~1, pp. 138--147, Feb. 2016.

\bibitem{byon2015adaptive}
E.~Byon, Y.~Choe, and N.~Yampikulsakul, ``Adaptive learning in time-variant
  processes with application to wind power systems,'' \emph{IEEE Trans. Autom.
  Sci. Eng}, vol.~13, no.~2, pp. 997--1007, Apr. 2016.

\bibitem{soman2010review}
S.~S. Soman, H.~Zareipour, O.~Malik, and P.~Mandal, ``A review of wind power
  and wind speed forecasting methods with different time horizons,'' in
  \emph{North American Power Symposium (NAPS), 2010}.\hskip 1em plus 0.5em
  minus 0.4em\relax IEEE, 2010, pp. 1--8.

\bibitem{HG2010_PCE}
A.~S. Hering and M.~G. Genton, ``Powering up with space-time wind
  forecasting,'' \emph{Journal of the American Statistical Association}, vol.
  105, no. 489, pp. 92--104, 2010.

\bibitem{Bouffard2008}
F.~Bouffard and F.~D. Galiana, ``Stochastic security for operations planning
  with significant wind power generation,'' \emph{IEEE Trans. Power Syst.},
  vol.~23, no.~2, pp. 306 -- 316, May 2008.

\end{thebibliography}

\end{document}